\documentclass{aa501}
\usepackage{graphicx}
\begin{document}
   \title{Global physical conditions 
   of the interstellar medium \\
   in nearby galaxies
   \thanks
{Based on observations with ISO, an ESA project with instruments funded by
ESA Member States (especially the PI countries: France, Germany, the
Netherlands and the United Kingdom) and with the participation of ISAS and
NASA.}}

   \author{T. Negishi\inst{1}
           \thanks{\emph{Present address:}
           NIKON Co., 1-10-1 Azamizodai, Sagamihara, Kanagawa 228-0828, Japan}
           \and T. Onaka\inst{1}
           \and K.-W. Chan\inst{1}
           \and T. L. Roellig\inst{2}
          }

   \offprints{T. Onaka (onaka@astron.s.u-tokyo.ac.jp)}

   \institute{Department of Astronomy, School of Science, University of 
Tokyo, 7-3-1 Hongo, Bunkyo-ku, Tokyo 113-0033, Japan
             \and  MS245-6, NASA Ames Research Center, Moffett Field,
             CA94035-1000, U.S.A.\\
             }

   \date{the date of receipt and acceptance should be inserted later
   }

\abstract{
Far-infrared spectra (43-197\,$\mu$m) of 34 nearby galaxies obtained by
the Long Wavelength Spectrometer (LWS) aboard the Infrared Space Observatory 
(ISO) were analyzed to investigate the general properties of interstellar 
matter 
in galaxies. The present sample includes not only normal galaxies but also 
starbursts and active galactic nuclei (AGNs).
Far-infrared forbidden lines, such as [\ion{C}{ii}]158\,$\mu$m, 
[\ion{O}{i}]63\,$\mu$m, [\ion{N}{ii}]122\,$\mu$m, and 
[\ion{O}{iii}]88\,$\mu$m, 
were detected in most of the sample galaxies. 
[\ion{O}{i}]145\,$\mu$m line was detected in 13 galaxies.
The line fluxes of [\ion{C}{ii}]158\,$\mu$m and [\ion{N}{ii}]122\,$\mu$m 
relative 
to the total far-infrared flux ($FIR$) decrease as the far-infrared color 
becomes 
bluer, while the ratio of the [\ion{O}{i}]63\,$\mu$m  
flux to $FIR$ does not show a systematic trend with 
the color.  The [\ion{O}{iii}]88\,$\mu$m to $FIR$ ratio shows a large 
scatter with a weak trend of increase with the color.
AGNs do not show any distinguishable trend from normal and starburst 
galaxies in 
the far-infrared spectra, suggesting that the far-infrared emission is 
mainly driven by star-formation activities even in AGNs.
We estimate the physical conditions of photodissociation regions (PDRs) in 
the sample galaxies, such as the far-ultraviolet radiation field intensity 
$G_\mathrm{0}$ and the gas density $n$ 
by assuming that all the observed [\ion{O}{i}]63\,$\mu$m and far-infrared 
continuum emissions come from PDRs. 
Comparison with PDR models indicates that $G_\mathrm{0}$ 
ranges from $10^2 -10^4$ and $n \sim 10^2 - 10^4 \mathrm{cm}^{-3}$. The 
present results also suggest 
that $n$ varies proportionally with $G_\mathrm{0}$.
The ratio of [\ion{C}{ii}]\,158$\mu$m to CO ($J=1-0$) line emission supports 
the linear increase in $n$ with $G_\mathrm{0}$.
We estimate that about a half of [\ion{C}{ii}]158\,$\mu$m 
emission originates from PDRs and
attribute the rest to the emission as coming from low-density diffuse 
ionized gas.
The estimated intensity of [\ion{C}{ii}]158\,$\mu$m from the ionized gas
is compatible with the observed intensity of 
[\ion{N}{ii}]122\,$\mu$m if both lines come from the same diffuse ionized 
gas.  The present analysis suggests that the decrease in 
[\ion{C}{ii}]158\,$\mu$m$/FIR$ with the 
far-infrared color may not be accounted for by the decrease in the 
photoelectric heating efficiency owing to the increase in positive charges 
of dust grains
because a measure of the efficiency, $G_\mathrm{0}/n$, is found to stay 
constant with the far-infrared color.  Instead the decrease
can be interpreted in terms of either the increase in the collisional 
de-excitation of 
the [\ion{C}{ii}] transition due to the increase in the gas density or the 
decrease in the 
ionized component relative to the far-infrared intensity suggested by the 
decrease in [\ion{N}{ii}]122\,$\mu$m$/FIR$.
Based on the present analysis, we derive average relations of the 
far-infrared 
color with $G_\mathrm{0}$ and $n$ in galaxies, which can be applied to the 
investigation 
of 
interstellar matter in distant galaxies.
      \keywords{galaxies: ISM -- infrared: ISM -- lines and bands: ISM -- 
	  radiation mechanisms: thermal
               }
}
   \titlerunning{Interstellar medium in nearby galaxies}
   \authorrunning{T. Negishi et al.}
   
      \maketitle

%

\section{Introduction}

Mechanisms for heating interstellar gas include collisions, radiation from 
stars, shocks, and cosmic rays.
Examination of the spectral lines that cool the gas can help determine the 
dominant excitation mechanisms and conditions. 
Studies of particular regions in our Galaxy and observations of 
external galaxies have suggested that stellar ultraviolet radiation can 
ionize 
vast volumes of a galaxy and that the far-ultraviolet (FUV) radiation 
impinging 
on neutral cloud surfaces is responsible for a large fraction of the observed 
far-infrared (FIR) spectral line emission that cools the gas 
(Crawford et al. \cite {Crawford}; Stacey et al. \cite {Stacey}; 
Shibai et al. \cite {Shibai}).
Tielens \& Hollenbach (\cite{TH85}) defined photodissociation regions (PDRs) 
as ``a 
neutral region where FUV radiation dominates the heating and/or some 
important aspect of the chemistry". 
Thus PDRs include most of the atomic gas in a galaxy, both in diffuse clouds 
and in dense regions (for a recent review, 
Hollenbach \& Tielens \cite{Hollenbach}). 

[\ion{C}{ii}]158\,$\mu$m and [\ion{O}{i}]63\,$\mu$m lines are important 
coolants 
in PDRs, while gas heating in PDRs is thought to be dominated by energetic 
photoelectrons ejected from dust grains following FUV photon absorption 
(Watson \cite{Watson}; de Jong \cite{de Jong}).
For galactic nuclei and star-forming regions in the spiral arms, 
most of the observed [\ion{C}{ii}] line emission arises from PDRs 
on molecular cloud surfaces. 
However, integrated over the disks of spiral galaxies, a substantial 
fraction 
may also arise from ``standard" atomic clouds, i.e., the cold neutral 
medium gas 
regions (CNM) (Madden et al. \cite {Madden}; Bennett et al. \cite{Bennett}) 
or from extended low-density warm ionized gas 
regions (ELDWIM) (Heiles \cite {Heiles}).
Contributions from the various gas phases can be estimated by observations 
of several FIR forbidden lines (Carral et al. \cite{Carral}; Luhman et al. 
\cite{Luhman}).  

With the LWS on board the ISO (Clegg et al. \cite{Clegg}) 
it now becomes possible to measure far-infrared lines not only from 
infrared-bright galaxies 
but also from normal galaxies (for latest reviews, Genzel \& Cesarsky 
\cite{Genzel}; Fischer \cite{Fischer}).  Lord et al. (\cite {Lord}) 
made observations of several FIR fine-structure forbidden lines
in a normal galaxy NGC5713 and found a fairly high [\ion{C}{ii}]158\,$\mu$m
to the FIR intensity ratio.
Smith \& Madden (\cite{Smith}) made observations of [\ion{N}{ii}]122\,$\mu$m 
and
[\ion{C}{ii}]158\,$\mu$m lines for five spiral galaxies in the Virgo cluster 
and found
enhanced ratios of [\ion{C}{ii}] to CO($J=1-0$) intensities in two of the 
five galaxies.
They interpreted the results in terms of either low metallicity or an 
increase in the contribution from the CNM.
Braine \& Hughes (\cite{Braine}) obtained a complete FIR spectrum of
a normal disk galaxy NGC4414 and investigated the physical conditions of the 
interstellar medium.
Leech et al. (\cite{Leech}) presented observations of 
[\ion{C}{ii}]158\,$\mu$m in 19
quiescent galaxies in the Virgo cluster.  They indicated a good correlation 
between [\ion{C}{ii}] and far-infrared intensities and a trend of increasing 
[\ion{C}{ii}]-to-far-infrared
intensity ratio with increasing galaxy lateness, which has been shown to be 
related to the star-formation rate (Pierini et al. \cite{Pierini}).
Malhotra et al. (\cite{Malhotra2001a}) reported observations
of 4 early-type galaxies with ISO/LWS and interpreted the observed low ratio 
of [\ion{C}{ii}]158\,$\mu$m to far-infrared intensities in terms of the soft 
radiation
field in the target galaxies.  Hunter et al. ({\cite{Hunter}) presented 
observations of 5 irregular galaxies with ISO/LWS and found  
strong [\ion{C}{ii}]158\,$\mu$m emission relative to the far-infrared 
continuum.  

Malhotra et al. (\cite{Malhotra1997}, \cite{Malhotra2001}) have investigated 
the far-infrared properties of 60 
nearby normal galaxies based on line-spectroscopic observations mainly of  
[\ion{C}{ii}]158\,$\mu$m and [\ion{O}{i}]63\,$\mu$m line emissions. They 
complemented their 
line data with IRAS photometry to estimate the far-infrared 
continuum intensity $FIR$ (Helou et al. \cite{Helou}) and
found a trend that the ratio of the [\ion{C}{ii}] line intensity to 
$FIR$ decreases with the far-infrared color becoming bluer.
Several interpretations have been proposed for the trend (Malhotra et al. 
\cite{Malhotra1997}; 
Genzel \& Cesarsky \cite{Genzel}; Helou et al. \cite{Helou2001}).
Malhotra et al. (\cite{Malhotra2001}) favor the 
interpretation of the decrease in the photoelectron yield owing to the
increase in positive charges 
of dust grains under strong ultraviolet radiation.

The FIR continuum emission shorter than 60\,$\mu$m is dominated by the 
emission 
from stochastically heated very small grains (Desert et al. \cite{Desert}; 
Dwek et al. 
 \cite{Dwek1997}; Onaka \cite{Onaka}; Dale et al. \cite{Dale}; Li \& Draine 
\cite{Li}).
The spectral energy distribution (SED) longer than 100\,$\mu$m is crucial to 
correctly estimate the thermal emission from submicron dust grains and 
understand the FIR SED of galaxies.
In this paper we investigate the FIR properties of nearby galaxies based on 
LWS full grating spectra from 43 to 197\,$\mu$m including both line and 
continuum emission.  The continuum spectra longer than 100\,$\mu$m enable 
better estimates of
the average temperature of submicron dust grains as well as the strength
of the interstellar radiation field. 
Because the aperture size of LWS is large ($\sim$ 80 \arcsec), the spectra 
of galaxies include contributions from various interstellar regions within
the galaxies.
This paper investigates the mechanism of gas heating with the aim of a 
better 
understanding of the global physical conditions of the interstellar medium 
in galaxies.

\begin{table*}[ht]
\caption{List of the galaxies in the present analysis}
\begin{center}
\begin{tabular}{cccccc} \hline
Galaxy & Type  & Morphology  & ISO(60)$^a$
[Jy] & R(60/100)$^b$ & visual size [\arcmin]\\\hline
Cen A & AGN  & S0 pec, Sy2 & 98.4  & 0.43 & $25.7 \times 20.0$ \\
Circinus & AGN  & SA(s)b:, Sy2 & 334.1  & 0.71  & $6.9 \times 3.0$\\
IC2554 & normal  & SB(s)bc pec: & 15.9  & 0.52  & $3.1 \times 1.3$\\
IRAS00506+7248 & normal &  & 24.1  & 0.66  \\
IRAS13242-5713 & normal & S... & 89.2  & 0.70  & $1.1: \times 0.2 $\\
M51 & normal  & SA(s)bc pec, HII, Sy2.5 & 36.4  & 0.38  & $11.2 \times 
6.9$\\
M82 & starburst & I0, Sbrst, HII & 1486.6  & 0.87 & $11.2 \times 4.3$  \\
M83 & starburst  & SAB(s)c, HII, Sbrst & 138.0  & 0.60 & $12.9 \times 11.5 
$\\
Maffei2 & normal  & SAB(rs)bc: & 94.8  & 0.47 & $5.82 \times  1.57$ \\
NGC1068 & AGN  & (R)SA(rs)b, Sy1, Sy2 & 206.3  & 0.62 & $7.1 \times   6.0$ 
\\
NGC1097 & starburst  & (R'$_1$:)SB(r'l)b, Sy1 & 49.6  & 0.54  & $9.3 \times   
6.3$ \\
NGC1365 & AGN  & (R')SBb(s)b, Sy1.8 & 92.0  & 0.52  & $11.2 \times 6.2$\\
NGC2146 & starburst  & SB(s)ab pec, HII & 163.3  & 0.65 & $6.0 \times   3.4 
$\\
NGC253 & starburst & SAB(s)c, HII, Sbrst & 1044.7  & 0.70 & $27.5  \times  
6.8$  
\\
NGC3256 & starburst  & Pec, merger, HII & 107.8  & 0.71  & $3.8 \times 
2.1$\\
NGC3690 & starburst & IBm pec, HII & 121.7  & 1.00 & $2.0 \times 1.5 $ \\
NGC4038 & starburst & SB(s)m pec & 21.5  & 0.52  & $5.2 \times 3.1 $\\
NGC4041 & normal  & SA(rs)bc: & 11.7  & 0.43 & $2.7 \times 2.5$ \\
NGC4414 & normal  & SA(rs)c? & 24.8  & 0.38 & $3.6 \times 2.0$ \\
NGC4945 & starburst  & SB(s)cd: sp, Sy2 & 577.2  & 0.50 & $20.0 \times 3.8 
$\\
NGC520 & starburst &  & 37.9  & 0.65  & $1.9  \times  0.7$ \\
NGC5430 & starburst & SB(s)b, HII, Sbrst & 9.2  & 0.51 & $2.2 \times   1.1 $ 
\\
NGC5937 & normal & (R')SAB(rs)b pec & 10.0  & 0.46 & $1.9 \times   1.1$  \\
NGC6156 & normal & (R'$_1$)SAB(rs)c & 20.9  & 0.61 &$ 1.6 \times   1.4 $ \\
NGC6240 & starburst  & I0: pec, LINER, Sy2 & 25.8  & 0.87 & $2.1 \times 1.1$ 
\\
NGC6764 & starburst  & SB(s)bc, LINER, Sy2 & 4.6  & 0.50 & $2.3 \times 1.3$ 
\\
NGC6810 & normal  & SA(s)ab:sp, Sy2 & 17.0  & 0.48  & $3.2 \times 0.9$\\
NGC6824 & normal  & SA(s)b: & 6.8  & 0.43 & $1.7 \times 1.2$ \\
NGC6946 & starburst & SAB(rs)cd, HII & 58.6  & 0.50 & $11.5 \times   9.8$  
\\
NGC7469 & starburst & (R')SAB(rs)a, Sy1.2 & 24.7  & 0.69  & $1.5 \times 1.1 
$\\
NGC7552 & starburst & (R')SB(s)ab, HII, LINER & 74.7  & 0.59 & $3.4 \times 
2.7$  
\\
NGC7582 & AGN& (R'$_1$)SB(s)ab, Sy2 & 51.9  & 0.64  & $5.0 \times 2.1$  \\
NGC7714 & starburst & SB(s)b:pec, HII, LINER & 12.7  & 0.97& $1.9 \times 
1.4$   
\\
NGC891 & normal  & SA(s)b? sp, HII & 24.3  & 0.31 & $13.5 \times 2.5$ 
\\\hline
\end{tabular}
\end{center}
$a$ Flux density at 60\,$\mu$m derived from the LWS spectra convolved with 
the 
IRAS band
filters.

$b$ Ratio of the 60\,$\mu$m and 100\,$\mu$m flux 
densities (in Jy) derived from the LWS spectra with the IRAS band filters.
\label{galaxiesData}
\end{table*}

\section{Data reduction and observational results}

Full grating scan spectra of $43 - 197$\,$\mu$m were obtained for 9 galaxies 
with the LWS01 mode in the open time programs of TONAKA.IRGAL and GALIR.
In addition, we extracted LWS01 full grating scan data of 25 galaxies from 
the ISO archival database for 
a total of 34 nearby galaxies LWS spectra that were analyzed in the present 
study.
The sample includes various types of galaxies, ranging from active galactic
nuclei (AGNs), starburst, 
to normal galaxies.  Table~\ref{galaxiesData} lists the present sample, 
where the flux density at 60\,$\mu$m and the FIR color $R(60/100)$ are
derived from the LWS spectra convolved with the IRAS band filters
(for the data reduction, see below).  
We excluded Arp220 and II ZW40 from the present analysis, both of which 
have also been observed with LWS01.
The former shows a spectrum optically thick even in the far-infrared
(Fischer \cite{Fischer}; Malhotra et al. \cite{Malhotra2001}), for 
which the present simple analysis cannot be applied, 
and the continuum of the latter object is faint and suffered from the 
uncertainty in the dark current estimate.
The spectral resolution ($\lambda/\Delta\lambda$) of LW01 mode
was about 200 and the beam size of LWS
was estimated to be $60\arcsec - 80\arcsec$ (Gry \cite{gry}).
The sample galaxies have optical sizes of $1\arcmin -  30\arcmin$ 
and thus for some galaxies only the central portion was included in the LWS 
beam.
Smith \& Harvey (\cite{Smith&Harvey}) reported observations of far-infrared 
emitting regions in 
external galaxies from the Kuiper Airborne Observatory (KAO), indicating that 
most 
far-infrared emission is concentrated in the central $30\arcsec$ regions of 
the galaxies.
LWS observations are supposed to detect most of the far-infrared 
emission and therefore probe
the properties of central part of the galaxies.

In the present study, we used the Standard Processed Data (SPD) of 
off-line processing (OLP) version 9 products provided by the ISO data center. 
The dark current and the drift in the detector responsivity were corrected 
by using the LWS Interactive Analysis software (LIA version 7.3)\footnote 
{LIA is a joint development of the ISO-LWS instrument team at RAL, UK, the 
PI institute and IPAC.}.
The ISO Spectral Analysis Package (ISAP version 1.6a)\footnote
{The ISO Spectral Analysis Package (ISAP) is a joint development by the LWS 
and SWS Instrument Teams and Data Centers. Contributing institutes are CESR, 
IAS, IPAC, MPE, RAL and SRON.} 
was then used for further data reduction. 
The continuum spectra were stitched together by shifting each detector 
signal with the offset method, adjusted to the SW5 channel in most galaxies.
In some cases where the SW5 channel is noisy, the adjustment was made 
referring to the LW3 or LW4 channels.
The offsets between the detectors were typically less than 20\%.
The line flux, the total far-infrared flux, and the dust temperature of the 
continuum emission were derived by ISAP.  The 60 and 100\,$\mu$m flux densities
from the LWS spectra were found to agree with the IRAS data within about 20\%.
The uncertainties in LWS spectra were suggested to be about $15-20$\% in 
previous
works (e.g., Braine and Hughes \cite{Braine}; Unger et al. \cite{Unger}) and we 
adopt 20\% errors for the flux uncertainty.

The continuum emission shorter than the 60\,$\mu$m region is affected by the
contribution from very small grains.
To derive a typical temperature of submicron grains $T_\mathrm{d}$ in each 
galaxy, 
we fitted the LWS spectrum for $\lambda \ge$ 80\,$\mu$m with the following 
equation:

\begin{equation}
\label{fittingFunction}
F(\lambda) = \Omega\; \tau_{0.55} \left(\frac{0.55 \mu m}{\lambda} 
\right) B_{\lambda}(T_\mathrm{d}),
\end{equation}

\vspace{0.5cm}
\resizebox{16.5cm}{!}{\includegraphics[angle=90]{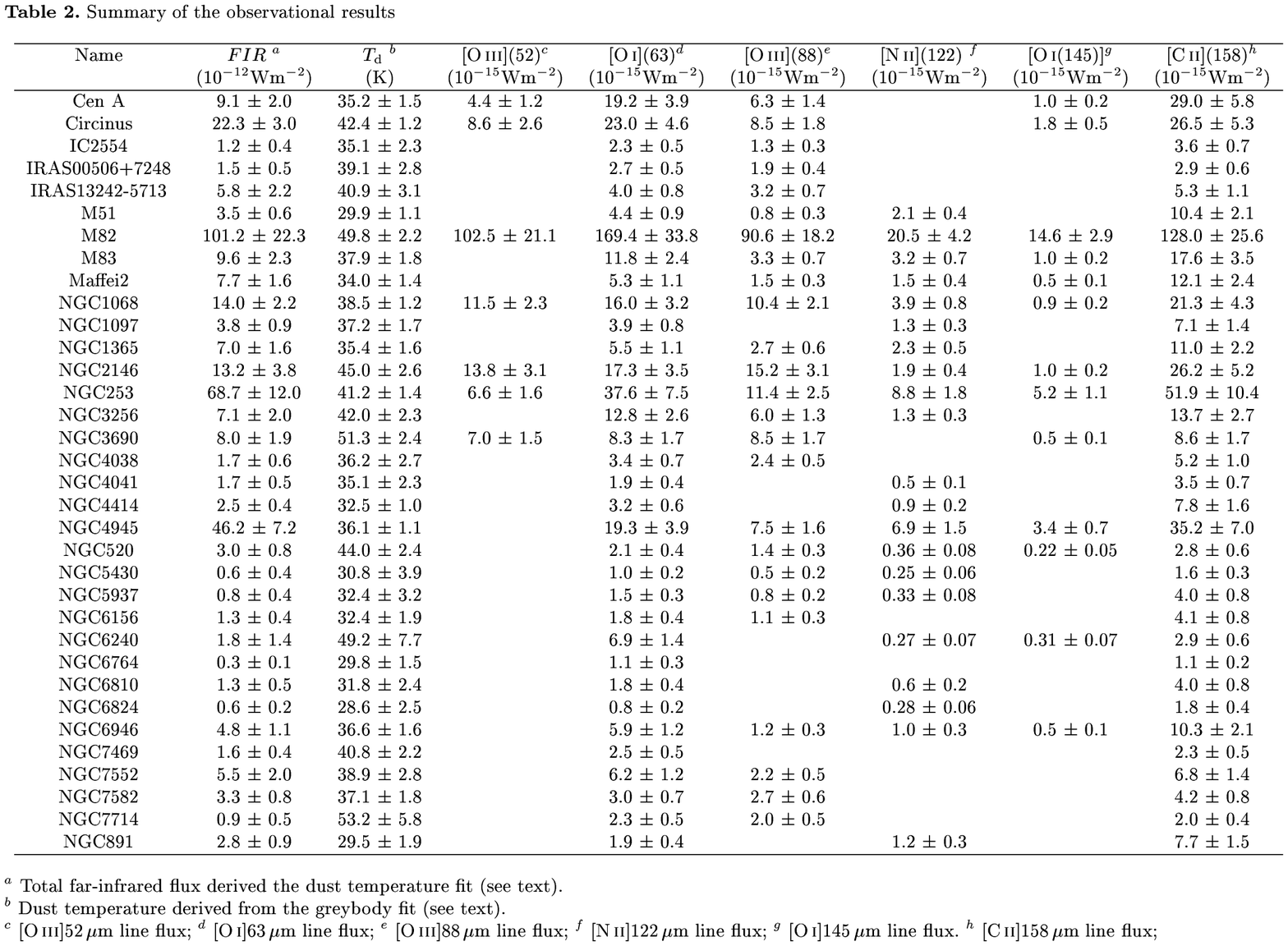}}
\label{tab2}
\clearpage
\setcounter{table}{2}

\noindent 
where $\Omega$, $\tau _{0.55}$, and $B_{\lambda}(T_\mathrm{d})$ are the 
solid 
angle of the object, the optical depth at 0.55\,$\mu$m, and the Planck 
function of temperature $T_\mathrm{d}$, respectively. 
The parameters $\Omega$ and $\tau _\mathrm{0.55}$ cannot be determined 
independently and only the product of them is a meaningful parameter.
We assumed that the dust emissivity is proportional to 1/$\lambda$, which 
provides reasonably good fits for the present sample of galaxies.
A typical example of the fit is shown in Fig.~\ref{SED}.
We then define the total infrared integrated flux from submicron dust grains 
as

\begin{equation}
FIR = \int_{0}^{\infty} F(\lambda)d\lambda.
\end{equation}
Note that $FIR$ does not include the excess emission shorter than 
80\,$\mu$m, 
which is attributed to the emission from very small dust grains.
The uncertainties in the flux level of the SW1 detector due to the memory 
effect and the spectral shape of the excess emission
make it difficult to estimate the excess emission accurately.
The excess emission may be about 30\% of $FIR$ according to the study of the 
diffuse Galactic emission of Dwek et al. (\cite{Dwek1997}).  The present $FIR$
is larger by 30\% on average than the far-infrared flux for $42-122$\,$\mu$m
estimated from the IRAS 60 and 100\,$\mu$m flux densities (Helou et al.
\cite{Helou}).

\begin{figure}
\resizebox{\hsize}{!}{\includegraphics{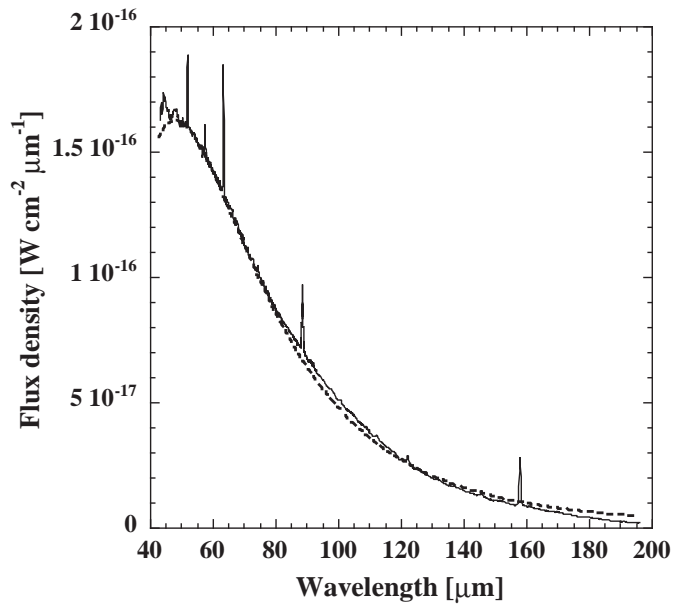}}
\caption{
The LWS spectrum and the fitted curve (Eq.~(\ref{fittingFunction})) for 
M82. 
}
\label{SED}
\end{figure}

The observational results are summarized in Table~\ref{tab2}. 
In most of the sample galaxies, far-infrared forbidden lines, such as 
[\ion{O}{i}]63\,$\mu$m, [\ion{O}{iii}]88\,$\mu$m, [\ion{N}{ii}]122\,$\mu$m, 
and [\ion{C}{ii}]158\,$\mu$m, were detected.
[\ion{O}{i}]145\,$\mu$m was detected only in 13 galaxies, while
only 8 galaxies show detectable [\ion{O}{iii}]52\,$\mu$m emission.
The indicated errors include the relative errors estimated from the uncertainty
in the base line fit and 20\% flux uncertainty.  The error in $FIR$ is estimated
from the uncertainty in the temperature determination, which mostly
comes from the goodness of the fit.

[\ion{C}{ii}]158\,$\mu$m is one of the most important lines for the 
diagnosis of physical conditions of PDRs because of its 
large luminosity and low critical density for collisional excitation.
However, the carbon atom has a lower ionization energy (11.26eV) than hydrogen 
(13.6eV), 
and carbon ions are expected to be present not only in the neutral region, 
such as PDRs, but also in the ionized regions.

[\ion{N}{ii}]122\,$\mu$m is a good tracer of diffuse 
low-density ionized gas, such as ELDWIM, because the ionization energy of 
nitrogen atom nearly equals to that of hydrogen atom and the critical 
density 
for collisional excitation by electrons is about 300 electrons cm$^{-3}$.

Oxygen atoms have an ionization energy of 13.6eV almost the same as that of 
hydrogen.  [\ion{O}{i}]63\,$\mu$m is one of the most luminous lines as well 
as  
[\ion{C}{ii}]158\,$\mu$m and it becomes a more efficient cooling line than 
[\ion{C}{ii}]158\,$\mu$m in high-density gases.
Together with the upper level transition at 145\,$\mu$m, it is an important
probe for neutral gas.  [\ion{O}{i}]145\,$\mu$m was weak and detected 
only in a limited number of galaxies, and we 
cannot examine the major fraction of the sample galaxies by using the
[\ion{O}{i}]145\,$\mu$m line.
[\ion{O}{iii}]88\,$\mu$m is a luminous line of dense ionized gas.
It has an upper level transition at 52\,$\mu$m and
the line ratio of the 52\,$\mu$m to 88\,$\mu$m emission can be used to 
derive the electron density of the ionized gas (e.g., Moorwood et al. 
\cite{Moorwood}).
Unfortunately, the spectra in the 52\,$\mu$m region do not have a 
sufficient 
signal to noise ratio to derive a reliable [\ion{O}{iii}]52\,$\mu$m line 
intensity for most of the present sample galaxies.  

In Table~\ref{tab3} we
list the electron density estimated from the ratio of the [\ion{O}{iii}] lines
and the neutral hydrogen density estimated from that of the [\ion{O}{i}] lines
for the galaxies in which [OIII]52\,$\mu$m emission was detected.
These are rough estimates and
should be taken with caution because of the large errors
in the obtained line ratios.  For about a half of the galaxies with the 
detected [\ion{O}{iii}]52\,$\mu$m emission,
the line ratio is near the low-density limit and only upper limits of the
electron density are given.  In the derivation of the neutral hydrogen
density we assume
that the gas temperature is 1000K.  Even with this temperature the line
ratios are in the low-density limit for the galaxies listed in 
Table~\ref{tab3}.  For lower
temperatures the upper limits will further be decreased.

\begin{table}[ht]
\begin{center}
\caption{Densities estimated from the line ratios}
\label{tab3}
\begin{tabular}{ccccc}
\hline
object&$n_\mathrm{e}$ (cm$^{-3}$)$^a$&$n$ (cm$^{-3}$)$^b$\\
\hline
Cen A    & $ < 140$   & $< 56000$\\
Circinus & $ 170 \pm 130$ & $< 10000$\\
M82      & $ 200 \pm 120$ & $ < 1000$ \\
NGC1068  & $ 200 \pm 120$ & $ < 48000$ \\
NGC2146  & $ 125 \pm 100$ & $ < 45000$ \\
NGC253   & $ < 70$        & $-^c$ \\
NGC3690  & $ < 180$       & $ < 38000$ \\
\hline
\end{tabular}
\end{center}

$^a$ The electron density derived from the [\ion{O}{iii}]52\,$\mu$m to
88\,$\mu$m line ratio.

$^b$ The neutral hydrogen density derived from the [\ion{O}{i}]145\,$\mu$m to
63\,$\mu$m line ratio for the gas temperature of 1000K.

$^c$ The line ratio of [\ion{O}{i}] is too large and
no reasonable density can be derived for NGC253 (see Fig.~\ref{145/63} and
next section).
\end{table}

In Fig.~\ref{obsresult}a, the ratios of [\ion{C}{ii}]158\,$\mu$m and 
[\ion{N}{ii}]122\,$\mu$m fluxes to the far-infrared flux, 
[\ion{C}{ii}]$/FIR$ 
and [\ion{N}{ii}]$/FIR$, are plotted against the far-infrared color 
$R(60/100)$ 
while the ratio [\ion{O}{i}]63\,$\mu$m and [\ion{O}{iii}]88\,$\mu$m to 
$FIR$, [\ion{O}{i}]$/FIR$ and [\ion{O}{iii}]$/FIR$, are plotted 
against $R(60/100)$ in Fig.~\ref{obsresult}b.
Figure \ref{obsresult}a indicates a trend that
the ratio [\ion{C}{ii}]$/FIR$ decreases with $R(60/100)$.  The 
ratio [\ion{N}{ii}]$/FIR$ also seems to decrease with $R(60/100)$.  
On the other hand, [\ion{O}{i}]$/FIR$ does not show a clear systematic 
trend with $R(60/100)$.  There seems a weak trend that  
[\ion{O}{iii}]$/FIR$ increases with the color, though the scatter is quite 
large.

Similar trends in the ratios of the line intensities to the far-infrared 
intensity have been obtained for the normal galaxy sample (Malhotra et al. 
\cite{Malhotra1997}, \cite {Malhotra2001}).  
The present sample includes not only normal galaxies but also starburst 
galaxies and AGNs.  
In Fig.~\ref{Malhotra}, we plot the data for normal galaxies by Malhotra et 
al. (\cite{Malhotra2001}) (filled circles) together with the present sample 
(open squares) for comparison.
In the figure, the total far-infrared flux $FIR$[IRAS] is derived by the 
formula of Helou et al. (\cite{Helou}) based on the IRAS data to make direct 
comparison.  Both show a similar trend to each other except that
the data of normal galaxies seem to have a steeper trend with the color.
AGNs (e.g., Cen A, Circinus, NGC1068, and NGC7582) are located in the same 
trend as
normal and starburst galaxies, suggesting that the far-infrared emission of 
AGNs is driven mainly by star-formation activities.

\begin{figure}
\resizebox{\hsize}{!}{\includegraphics{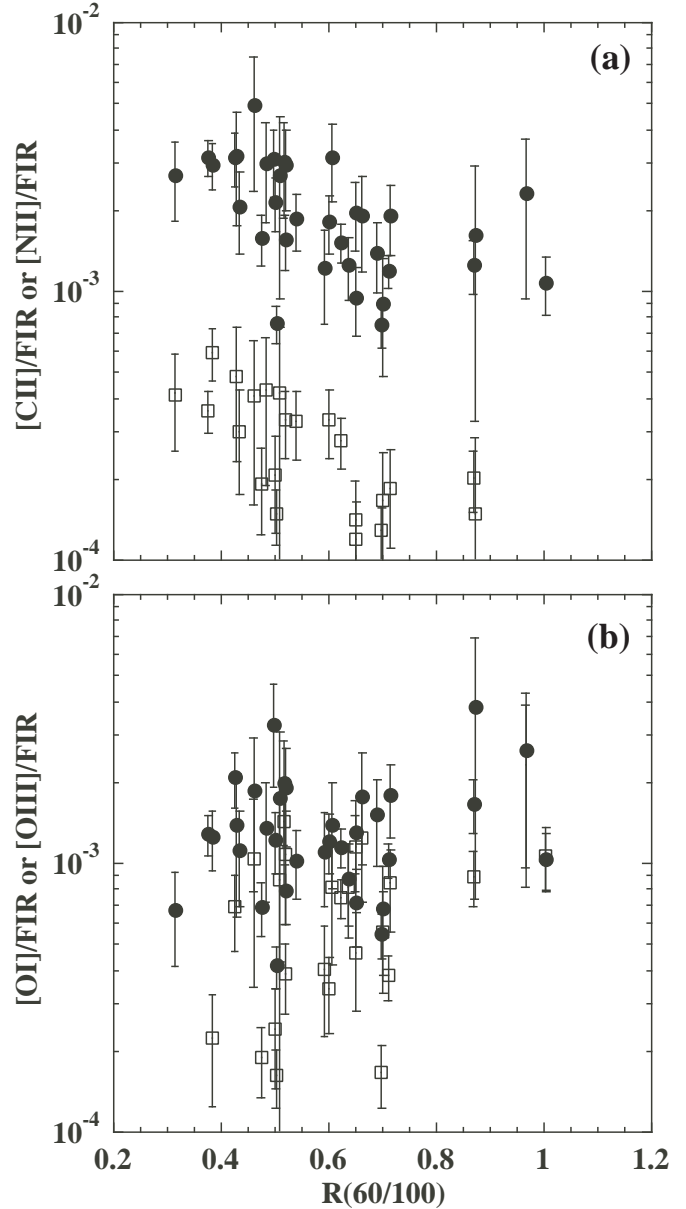}}
\caption{
{\bf a)} Ratios of the [\ion{C}{ii}]158\,$\mu$m flux and [\ion{N}{ii}]122\,$\mu$m 
flux to the total far-infrared flux $FIR$ against the far-infrared color 
$R(60/100)$.  Filled circles indicate [\ion{C}{ii}]$/FIR$ and open squares 
[\ion{N}{ii}]$/FIR$.
{\bf b)} Ratios of the [\ion{O}{i}]63\,$\mu$m flux and [\ion{O}{iii}]88\,$\mu$m 
flux 
to $FIR$ against $R(60/100)$. Filled circles indicate [\ion{O}{i}]$/FIR$ and 
open squares [\ion{O}{iii}]$/FIR$.
}
\label{obsresult}
\end{figure}

\begin{figure}
\resizebox{\hsize}{!}{\includegraphics{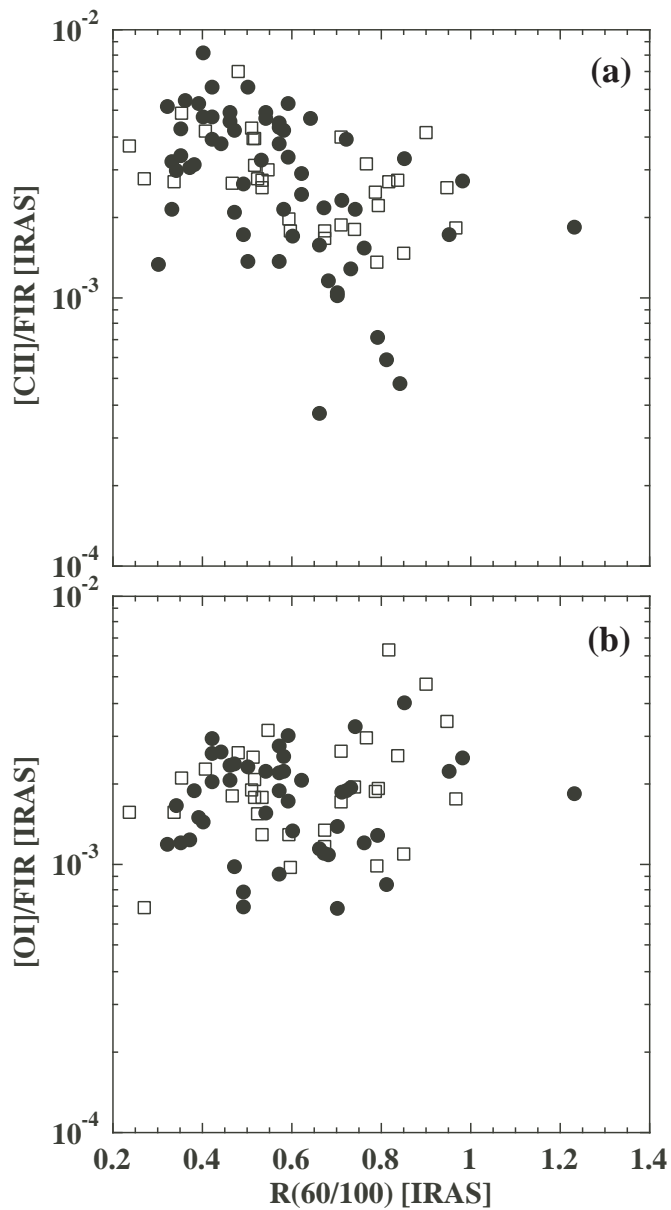}}
\caption{
{\bf a)} Ratio of the [\ion{C}{ii}]158\,$\mu$m line flux to the total 
far-infrared flux $FIR$ against the far-infrared $R(60/100)$. 
{\bf b)} Ratio of the [\ion{O}{i}]63\,$\mu$m line flux to $FIR$ against 
$R(60/100)$.  Filled circles are taken from 
Malhotra et al. (2000) and open squares indicate the present sample.
The value of $R(60/100)$ is based on IRAS data and 
$FIR$ is derived by the formula of Helou et al. (\cite{Helou}).
}
\label{Malhotra}
\end{figure}

\section{Physical conditions of interstellar medium}\label{sec3}

Together with the [\ion{O}{i}]63\,$\mu$m and 145\,$\mu$m lines, 
[\ion{C}{ii}]158\,$\mu$m can be used to derive
the physical conditions of the line-emitting regions based on PDR models 
(e.g., Tielens \& Hollenbach
 \cite {TH85}; Wolfire et al. \cite {WTH90}; Hollenbach \& Tielens 
\cite{Hollenbach}; Kaufman et al. \cite {Kaufman}), 
in which the major model parameters 
are the incident FUV radiation field flux $G_\mathrm{0}$ in units of the 
solar neighborhood 
value ($1.6 \times 10^{-6}$ W m$^{-2}$, Habing \cite{Habing}) 
and the neutral hydrogen gas density $n$.  
However, the [\ion{C}{ii}] line could also originate from ionized regions 
and the fraction of the contribution cannot be estimated {\it a 
priori}.  

To estimate the contribution to [\ion{C}{ii}]158\,$\mu$m from PDRs, we take 
two approaches similar to Malhotra et al. (\cite{Malhotra2001}).
First we assume that all the emission of [\ion{O}{i}]63\,$\mu$m and 
far-infrared 
continuum ($\lambda \ge 80\,\mu$m) comes from PDRs.  Since the temperature
of sub-micron sized dust grains is determined by the intensity of the incident
radiation (e.g., Onaka \cite{Onaka}), $G_\mathrm{0}$ can be estimated
from the dust temperature $T_\mathrm{d}$ derived by  Eq.~(\ref{fittingFunction}).
We used a semi-analytical equation of $G_\mathrm{0}$ and $T_\mathrm{d}$ 
given by 
Hollenbach et al. (\cite{HTT91}) with $A_\mathrm{V}$ = 0.5 because
a major fraction of [\ion{C}{ii}]158\,$\mu$m and [\ion{O}{i}]63\,$\mu$m 
emissions stem from the region of 
$A_\mathrm{V} \le 1$ (Kaufman et al. \cite{Kaufman}).
For M82 we derive $G_\mathrm{0} = 10^{3.4}$, while Kaufman et al. 
(\cite{Kaufman}) estimated $G_\mathrm{0} = 10^{3.5}$ by taking account
of several observed line intensities, suggesting that 
the present method provides a reasonable estimate of $G_\mathrm{0}$.

Then we compare the ratio of [\ion{O}{i}]$/FIR$ with the PDR model of 
Kaufman 
et al. (\cite{Kaufman}) with the derived $G_\mathrm{0}$ to estimate $n$.
Finally we estimate the intensity of [\ion{C}{ii}]158\,$\mu$m from PDRs with 
the derived $G_\mathrm{0}$ and $n$.
The current estimate of the flux uncertainty is 20\% and there 
may be an uncertainty in the PDR model due to the assumed geometry.
While $FIR$ may be underestimated by a few tens \% 
in the present analysis, 
it does not introduce a significant error compared to other uncertainties.

\begin{figure}
\resizebox{\hsize}{!}{\includegraphics{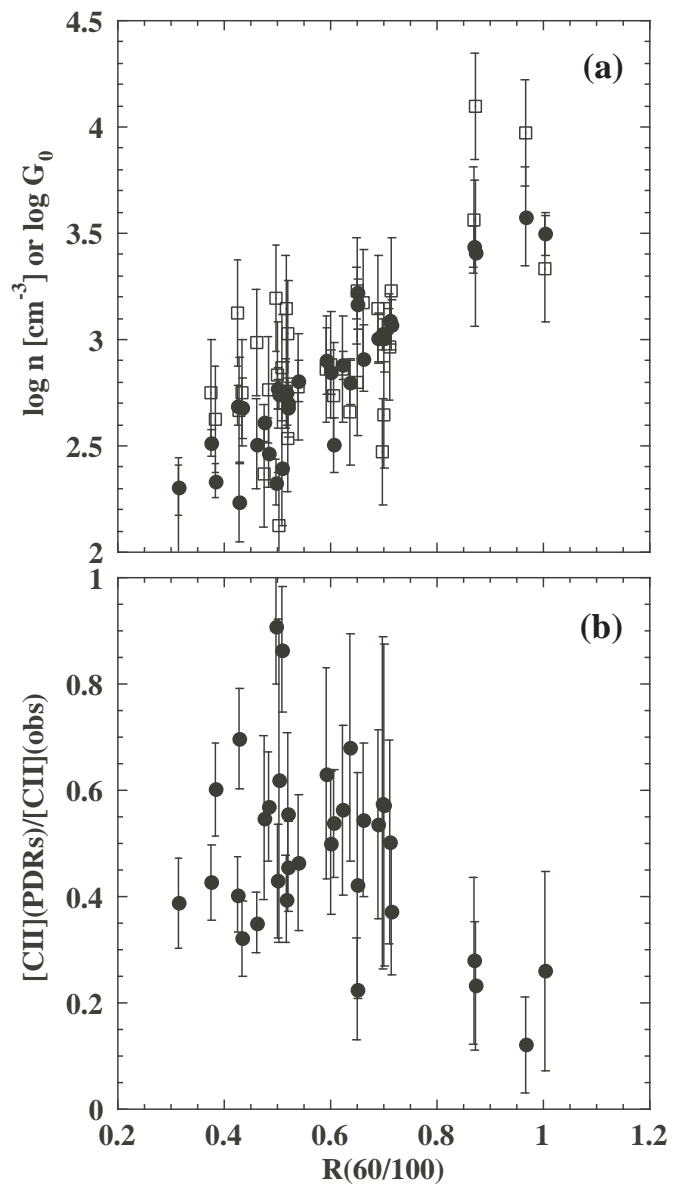}}
\caption{
{\bf a)}  FUV incident flux estimated by the dust temperature,  $G_\mathrm{0}$, 
and the derived cloud gas density, $n$, against the far-infrared color 
$R(60/100)$. 
Filled circles indicate $G_\mathrm{0}$ and open squares $n$ [cm$^{-3}$].
{\bf b)} Ratio of the [\ion{C}{ii}] intensity from PDRs to the observed total 
[\ion{C}{ii}] intensity vs. the $R(60/100)$.
}
\label{[OI]/FIRtest}
\end{figure}

Figure \ref{[OI]/FIRtest} shows $G_\mathrm{0}$ and $n$ against the color 
$R(60/100)$.  The error in $G_\mathrm{0}$ comes from the uncertainty in 
$T_d$, 
while that in $n$ is estimated from the uncertainty in [\ion{O}{i}]$/FIR$ to 
be about $\Delta \log n \simeq 0.3$ based on the PDR models.
Both parameters, $G_\mathrm{0}$ and $n$, increase clearly with
$R(60/100)$ and $n$ is found to be almost 
proportional to $G_\mathrm{0}$ ($\log n = (0.8 \pm 0.25) \log G_\mathrm{0}+ 
0.75_{-0.42}^{+0.26}$, see also Fig.~\ref{G0n}).  
The trend mainly comes from the observed constancy of [\ion{O}{i}]$/FIR$ 
against the color. 
In the PDR model the photoelectric heating efficiency is roughly a function 
of $G_\mathrm{0}/n$ (Bakes \& Tielens \cite{Bakes}).
For $G_\mathrm{0} \propto n$, the heating efficiency does not change and 
thus [\ion{O}{i}]$/FIR$ stays constant.
The [\ion{C}{ii}]158\,$\mu$m$/FIR$, on the other hand, decreases 
because the collisional de-excitation becomes efficient as $n$ increases:
the [\ion{C}{II}] line has a lower critical density for excitation 
($\sim 3 \times 10^3$\,cm$^{-3}$)
than [\ion{O}{i}]63\,$\mu$m ($\sim 1\times 10^{7}$\,cm$^{-3}$).
In this situation, we expect an increase in the gas temperature.
Because the gas temperature is already sufficiently high in the parameter 
range in question, the increase in the temperature does not affect
the line intensities appreciably.
Hence the decrease in [\ion{C}{ii}]$/FIR$ with color can be attributed 
to the increase in the collisional de-excitation of the [\ion{C}{ii}] 
transition for the present sample of the galaxies.    
The [\ion{C}{ii}]158\,$\mu$m emission from PDRs is estimated to be  $\sim 50 
\pm 30$\% of the total observed 
[\ion{C}{ii}]158\,$\mu$m emission in the present analysis and the
fraction of the PDR component is indicated to decrease with $R(60/100)$
(Fig.~\ref{[OI]/FIRtest}b).  Therefore, the decrease in the total 
[\ion{C}{ii}]
$/FIR$ is attributed to the decrease in the PDR component due to the 
thermalization of the [\ion{C}{ii}] transition.

In order to examine the reliability of the present data reduction and 
analysis we can compare the present results
with those of previous works for some individual galaxies.
The comparison is summarized in Table~\ref{ComparisonGalaxies}.
Colbert et al. (1999) and Unger et al. (\cite{Unger}) analyzed the same
LWS data and derived $G_\mathrm{0}$ and $n$ by 
using the PDR models of Kaufman et al. (1999) for M82 and Cen A, 
respectively.
The present analysis provides fairly good agreement with their results.
The line intensities derived in the present study also agree with those by 
Braine \& Hughes (\cite{Braine}) within the measurement errors.
Carral et al. (\cite{Carral}) reported the results of observations of NGC253 
and NGC3256 by the KAO
and obtained $G_\mathrm{0}$ and $n$ based on the PDR models of 
Wolfire et al. (\cite{WTH90}). While the line fluxes are in agreement 
with the present results
within the errors except for [\ion{O}{iii}]88\,$\mu$m of NGC253, the present
results indicate systematically low densities.  The difference can be 
attributed to the relatively
high gas temperature with the same $G_\mathrm{0}$ and $n$ in the models of 
Kaufman et al. because their models include the additional gas heating due to 
polycyclic aromatic hydrocarbons (PAHs).  
 
\begin{table*}[htb]
\begin{center}
\caption{Comparison with previous studies}
\label{ComparisonGalaxies}
\begin{tabular}{ccccccccc}
\hline
object&ref.$^a$&[\ion{O}{i}]63\,$\mu$m
&[\ion{O}{iii}]88\,$\mu$m
&[\ion{N}{ii}]122\,$\mu$m
&[\ion{O}{i}]145\,$\mu$m
&[\ion{C}{ii}]158\,$\mu$m
&$G_\mathrm{0}$
&$n$\\
&&(10$^{-15}$Wm$^{-2}$)
&(10$^{-15}$Wm$^{-2}$)
&(10$^{-15}$Wm$^{-2}$)
&(10$^{-15}$Wm$^{-2}$)
&(10$^{-15}$Wm$^{-2}$)
&
&(cm$^{-3}$)\\\hline
M82&p&169 $\pm$ 34&91 $\pm$ 18&21 $\pm$ 4&15 $\pm$ 3&128 $\pm$ 26&
10$^{3.4}$&10$^{3.6}$\\
&1& 176 $\pm$ 5 &86 $\pm$ 4&17 $\pm$ 3&12 $\pm$ 1&134 $\pm$ 1&
10$^{2.8}$&10$^{3.3}$\\

Cen A&p&19 $\pm$ 4&6.3 $\pm$ 1.4& &1.0 $\pm$ 0.2&29 $\pm$ 6&
10$^{2.7}$&10$^{3.1}$\\
&2&19.6&7.0&1.5&1.1&29.1&$\sim 10^{2}$&$\sim 10^{3}$\\

NGC4414&p&3.2 $\pm$ 0.6 &  & 0.9 $\pm$ 0.2& &7.8 $\pm$ 1.6\\
&3&3.3 $\pm$ 1&1 $\pm$ 0.5&1.3 $\pm$ 0.4&  &10.6 $\pm$ 2\\

NGC253&p&38 $\pm$ 7.5&11.4 $\pm$ 2.5&&&52 $\pm$ 10&10$^{3}$&10$^{2.5}$\\
&4&45 $\pm$ 6&6 $\pm$ 1&&&48 $\pm$ 2&10$^{4.3}$&10$^{4}$\\

NGC3256&p&12.8 $\pm$ 2.6&6.0 $\pm$ 1.3 &&&13.7 $\pm$ 2.7 &10$^{3.1}$&
10$^{3.2}$\\
&4&14.3 $\pm$ 2.6&4.8 $\pm$ 1.2&&&11.7 $\pm$ 2.5&10$^{3}$&10$^{3.9}$\\

\hline
\end{tabular}
\end{center}
$^a$ references: p present work; 1 Colbert et al. (\cite{Colbert}); 2
Unger et al. (\cite{Unger}); 3 Braine \& Hughes (\cite{Braine}); 
4 Carral et al. (\cite{Carral})
\end{table*}

The [\ion{C}{ii}] emission other than the PDR origin may be ascribed to the 
extended low density warm ionized medium (ELDWIM).  A large fraction of
[\ion{N}{ii}]122\,$\mu$m line emission is thought to stem mostly from the 
ELDWIM (Wright et al. \cite{Wright}; Heiles \cite 
{Heiles}; Bennett et al. \cite {Bennett}; Petuchowski et al. \cite 
{Petuchowski}). 
Figure \ref{NII/CII} plots the ratio of the observed [\ion{N}{ii}]122\,$\mu$m 
intensity to the non-PDR component of [\ion{C}{ii}]158\,$\mu$m.
The ratio shows a large scatter with a weak trend that the ratio 
decreases with $R(60/100)$.  Because of the uncertainties the trend may be 
spurious (see below).

All the data points are located within the range $0.1 - 0.7$ in 
Fig.~\ref{NII/CII}.  The line ratio of 
[\ion{N}{ii}]/[\ion{C}{ii}] in the ionized gas depends on the electron 
density, but is insensitive to the temperature of the ionized gas. To estimate 
the line ratio expected from ionized gas
we assume $T_\mathrm{e}=7000$K and the abundance of [\element[+][]{C}]/[H$^+$] = 
$4 \times 10^{-4} 
\delta_{\mathrm{C}^+}$ and [\element[+][]{N}]/[H$^+$] = $1 \times 10^{-4} 
\delta_{\mathrm{N}^+}$ with the depletion
factors as $\delta_{\mathrm{C}^+} = 0.65$ and $\delta_{\mathrm{N}^+} = 1$ 
in the following discussion
(Heiles \cite{Heiles}).  We adopted the collision coefficients for 
\element[+][]{N} from 
Stafford et al. (\cite{Stafford}) and those for \element[+][]{C} from Heiles 
(\cite{Heiles}).  
Recent {\it HST} observations indicate that the interstellar abundance of 
carbon and nitrogen in
the gas phase is fairly constant on various lines of sight in our Galaxy as 
$\delta_{\mathrm{C}} = 0.35 \pm 0.05$ (Sofia et al. \cite{sofia}) and 
$\delta_{\mathrm{N}} = 0.75 \pm 0.05$ (Meyer et al.
\cite{meyer}).  Based on these values the relative abundance of N to C 
will increase by 30\%.
The following discussion thus has an uncertainty of this level 
associated with the assumed abundance.

The lower boundary of the observed ratio 0.1 is then found to correspond to 
the low-density limit of the ratio
in the ionized gas.  The upper boundary 0.7 is obtained for a gas with
$n_\mathrm{e}$ = 120\,cm$^{-3}$.  
Petuchowski et al. (\cite {Petuchowski}) reported a large 
[\ion{N}{ii}]122\,$\mu$m
to 205\,$\mu$m line ratio in the central 850pc of M82 compared to the 
average ratio of the Milky Way (Wright et al. \cite{Wright}), 
indicating that a fair fraction of the [\ion{N}{ii}] line emission
comes from the ionized gas of $n_\mathrm{e} = 150-180$\,cm$^{-3}$ in M82.
The ratio of [\ion{N}{ii}]122\,$\mu$m to [\ion{C}{ii}]158\,$\mu$m of non-PDR 
origin for M82 
is about 0.3 in the present analysis, suggesting that there may be a 
significant contribution to the [CII] emission from
the low density diffuse ionized gas in the outer part ($>$ 850pc) of the 
galaxy.  The observed intensity
is compatible with the interpretation that
the non-PDR component of [\ion{C}{ii}]158\,$\mu$m comes from the
ionized gas that emits [\ion{N}{ii}]122\,$\mu$m.

\begin{figure}
\resizebox{\hsize}{!}{\includegraphics{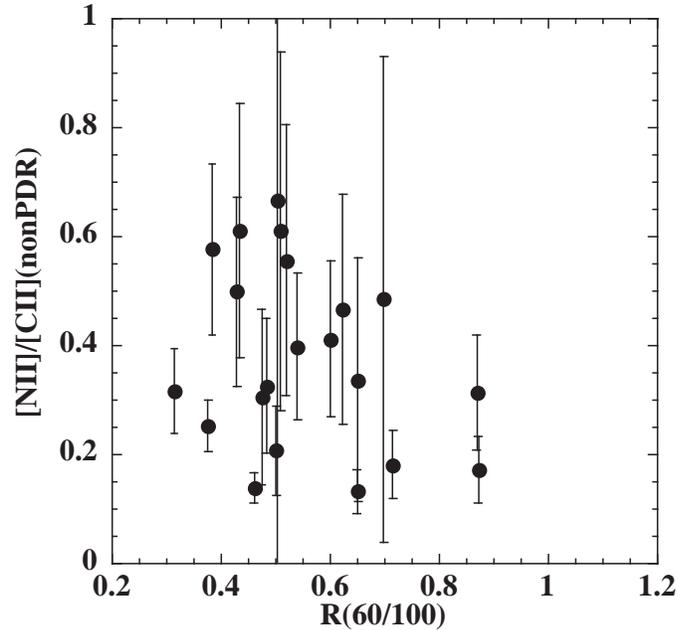}}
\caption{
	Ratio of [\ion{N}{ii}]122\,$\mu$m intensity to the non-PDR 
	component of [\ion{C}{ii}]158\,$\mu$m against $R(60/100)$. 
	}
\label{NII/CII}
\end{figure}

\begin{figure}
\resizebox{\hsize}{!}{\includegraphics{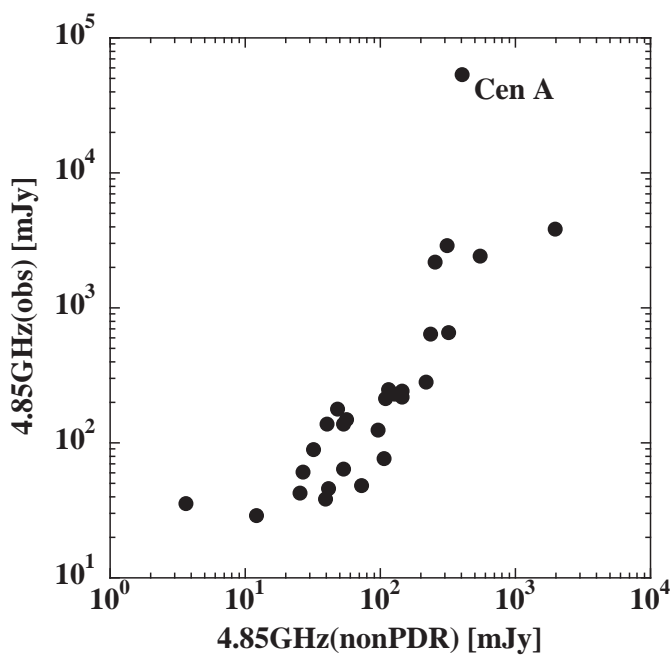}}
\caption{
	The observed 4.85GHz radio continuum flux density plotted against
	the 4.85GHz radio continuum flux density predicted 
	from the non-PDR component of [\ion{C}{ii}]158\,$\mu$m emission (see 
text).
	The observed radio flux densities are taken from Gregory et al. 
	(1991, 1994), Becker et al. (1991), and Griffith et al. (1994). 
	}
\label{radio_fig}
\end{figure}

The ionized gas also emits a radio continuum.  The intensity of free-free
transition is written for $T_\mathrm{e} = 7000 \mathrm{K}$ by

\begin{equation}
I(4.85\hbox{GHz})[\hbox{mJy sr}^{-1}] = 3.12 \times 10^{-14} n_\mathrm{e}^2 
l,
\label{radio}
\end{equation}
where $n_\mathrm{e}$ is the electron density in cm$^{-3}$ and $l$ is the 
path 
length in cm (Spitzer \cite{Spitzer}).  The intensity of the
[\ion{C}{ii}] line from the ionized gas
is given by

\begin{equation}
I_{\hbox{[\ion{C}{ii}]}} (\mathrm{ELDWIM})  = (1/4\pi) 
L(T_\mathrm{e})n(C^+)n_\mathrm{e} l, 
\label{CII_ELDWIM}
\end{equation}
where $L(T_\mathrm{e})$ is the cooling function of [\ion{C}{ii}]158\,$\mu$m 
(e.g., Hayes \& Nussbaumer \cite {H&N}). 
For a given [\ion{C}{ii}]158\,$\mu$m intensity, the corresponding radio 
continuum intensity increases with $n_\mathrm{e}$ as can
be estimated through Eqs.~(\ref{radio}) and (\ref{CII_ELDWIM}).
As a conservative upper limit we assume that the non-PDR component of 
[\ion{C}{ii}]
158\,$\mu$m emission comes from the ionized gas of density 200\,cm$^{-3}$. 
Figure \ref{radio_fig} shows the comparison between the predicted and observed 
flux densities.
We take the 4.85GHz radio continuum data from Becker et al. (\cite{Becker}), 
Gregory \& Condon (\cite{Gregory91}), Gregory et al. (\cite{Gregory94}), and 
Griffith et al. (\cite{Griffith}).  The beam size of the radio observations
was about $4\arcmin$ (FWHM).  
Observations of 1.49\,GHz and 1.425\,GHz indicate that most of the radio 
continuum 
emission comes from the area within about $1\arcmin$  of the center of the 
infrared 
emission (Condon et al. \cite{Condon90}, \cite{Condon96}).  We assume 
that the observed 4.85\,GHz emission also comes from the region within 
the LWS beam.  The predicted value shows a positive correlation with the 
observed flux density except for Cen A.  Cen A is a very strong radio source 
and most of the radio emission from Cen A is nonthermal 
(e.g. Sreekumar et al. \cite {Sreekumar}).
The observed radio intensities are larger than the upper limits predicted from  
[\ion{C}{ii}]158\,$\mu$m emission of the non-PDR origin for most of the sample
galaxies.  Thus the non-PDR component of 
[\ion{C}{ii}]158\,$\mu$m emission is compatible with the observed
radio continuum intensity
when it arises mostly from the low-density ionized gas for the present 
sample of galaxies.
\begin{figure}
\resizebox{\hsize}{!}{\includegraphics{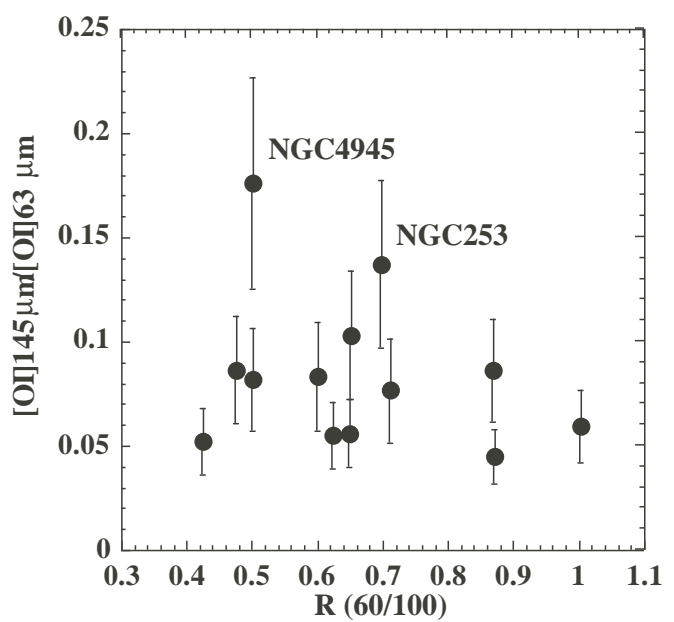}}
\caption{
Ratio of the intensity of the [\ion{O}{i}]145\,$\mu$m line to the 
[\ion{O}{i}]63\,$\mu$m line against $R(60/100)$.  
}
\label{145/63}
\end{figure}

Figure \ref{145/63} shows the ratio of [\ion{O}{i}]145\,$\mu$m to 
[\ion{O}{i}]63\,$\mu$m against
$R(60/100)$.  The observed ratio is in the range $0.05 - 
0.1$ for most galaxies, which is in agreement with the prediction of the
PDR model with the range 
of $G_\mathrm{0}$ and $n$ estimated in the present analysis.
The ratio does not exceed 0.1 within a reasonable range of the density and
temperature (Watson \cite{watson1984}).
However, a few galaxies clearly show higher ratios than the model.
They are NGC4945 and NGC253, both of which have large inclination angles 
(for NGC4945 and NGC253, $i \simeq 85\degr$ and $80\degr$, 
respectively).  NGC520, which shows the third largest ratio, has $i=60\degr$.
The absorption of [\ion{O}{i}]63\,$\mu$m in the interstellar 
medium may affect the ratio in these galaxies.  For a majority of the 
galaxies, however, this effect is probably not significant.

In the analysis described above it is difficult to properly
evaluate the uncertainties in the comparison with the model.
In order to examine how robust the derived conclusions are,
we take another approach to estimate the PDR contribution to the
[\ion{C}{ii}]158\,$\mu$m line emission.
Figure \ref{NII/CII} indicates that there seems no strong trend in the ratio 
of [\ion{N}{ii}]122\,$\mu$m to the non-PDR origin of 
[\ion{C}{ii}]158\,$\mu$m.
We thus simply assume that the contribution from the ionized gas to
[\ion{C}{ii}]158\,$\mu$m 
emission is proportional to the [\ion{N}{ii}]122\,$\mu$m intensity.
We take a mean value of Fig.~\ref{NII/CII} as

\begin{equation}
[\ion{C}{ii}]158\,\mu\mathrm{m}(\mathrm{ELDWIM}) = 3.5 \times 
[\ion{N}{ii}]122\,\mu\mathrm{m}.
\label{conversion}
\end{equation}
This relation corresponds to a gas of $n_\mathrm{e} = 35 \mathrm{cm}^{-3}$ 
for $T_\mathrm{e} = 7000$ K.
We estimate the intensity of [\ion{C}{ii}]158\,$\mu$m from the ionized gas 
based on the [\ion{N}{ii}]122\,$\mu$m intensity by 
using Eq.~(\ref{conversion}) and attribute the rest to that coming from 
PDRs.  We then
estimate $G_\mathrm{0}$ and $n$ from [\ion{O}{i}]/[\ion{C}{ii}] and 
([\ion{C}{ii}]+[\ion{O}{i}])$/FIR$.
Figure \ref{methodII}a shows $G_\mathrm{0}$ and $n$ estimated in this method.
The values of $G_\mathrm{0}$ derived in this method are in agreement with 
those obtained 
in the first approach within the estimated errors.  Thus the parameters 
derived in 
the second method are still compatible with the continuum spectrum of the 
LWS spectra.  We have obtained the same trend as in the first approach:
both $n$ and $G_\mathrm{0}$ increase with the color and $n$ is roughly
proportional to $G_\mathrm{0}$ (see also Fig.~\ref{G0n}).
We conclude that the linear increase of $n$ with
$G_\mathrm{0}$ is a rather secure result for the present sample galaxies.
The weak trend seen in Fig.~\ref{NII/CII} is not necessarily real.
Figure \ref{methodII}b shows the fraction of the PDR component of 
[\ion{C}{ii}] emission
derived in this analysis, suggesting that the PDR component does not vary 
with $R(60/100)$ in contrast to Fig.~\ref{[OI]/FIRtest}b.  Thus
in this analysis the decrease in [\ion{C}{ii}]$/FIR$ can be
interpreted in terms mainly of the decrease in the ionized component 
relative to $FIR$
as indicated in the decrease in [\ion{N}{ii}]$/FIR$ (Fig.~\ref{obsresult}a), 
though the decrease in [\ion{C}{ii}](PDR)$/FIR$ due to the 
thermalization also contributes partly.  

\begin{figure}
\resizebox{\hsize}{!}{\includegraphics{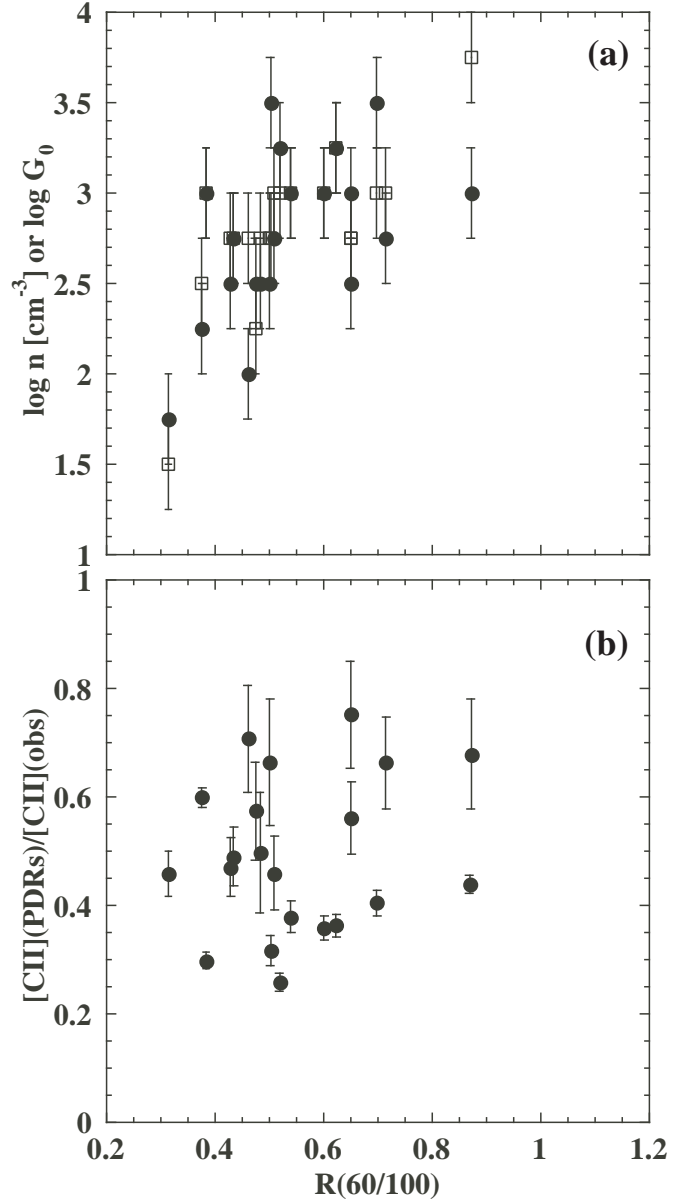}}
\caption{The results of the analysis based on the assumption
that the contribution from the ionized gas to 
[\ion{C}{ii}]158\,$\mu$m is proportional to [\ion{N}{ii}]122\,$\mu$m
as given by Eq.~(\ref{conversion}) (see text).
{\bf a)} FUV incident flux,  $G_\mathrm{0}$ (filled circle) and the gas density, 
$n$ (open square) derived from [\ion{O}{i}]/[\ion{C}{ii}] and 
([\ion{C}{ii}]+[\ion{O}{i}])$/FIR$ of the PDR model by Kaufman et al. (1999) 
against $R(60/100)$.  {\bf b)} The ratio of the estimated
[\ion{C}{ii}] intensity from PDRs to the observed total
[\ion{C}{ii}] intensity against $R(60/100)$.
}
\label{methodII}
\end{figure}

The ratio of [\ion{C}{ii}]158\,$\mu$m to \element[][12]{CO} (J=1-0) line 
intensity is another measure for the diagnosis of PDRs.
Because \element[+][]{C} is converted to CO as the gas shielding to prevent 
CO from 
dissociation becomes efficient, the thickness of \element[+][]{C} layer is a 
function of $G_\mathrm{0}/n$ (Mochizuki \& Onaka \cite{Mochizuki2001}).
The intensity ratio increases with $G_\mathrm{0}$ and decreases with $n$
and it is a function of $G_\mathrm{0}/n$ 
in the range of $G_\mathrm{0}$ and $n$ in question (Pierini et al. 
\cite{Pierini}; Kaufman et al. 
\cite{Kaufman}; Mochizuki \& Nakagawa \cite{Mochizuki}).
Metallicity also plays an important role in the \element[+][]{C} to CO 
conversion (Mochizuki et al. \cite{Mochizuki1994}).  Hence,
unless there is an appreciable variation in the metallicity in the sample 
galaxies, the ratio of 
[\ion{C}{ii}]/CO approximately varies with $G_\mathrm{0}/n$.
Figure \ref{CO} plots the ratio against the color.
The CO data are taken from Young et al. (\cite{Young}), Elfhag et al. 
(\cite{Elfhag}), Aalto et al. (\cite{Aalto}), Curran et al. (\cite{Curran}),
Stacey et al. (\cite{Stacey}), and Eckart et al. (\cite{Eckart}).  The CO 
observations had the beam size of $45-56$\arcsec.
We assume that the CO emission comes mostly from the central part of the 
galaxies and did not make any corrections for the beam size. 
Except for NGC6824, the ratio stays almost constant, supporting 
$n \propto G_\mathrm{0}$.  The constancy of $G_\mathrm{0}/n$ is also 
confirmed by the [\ion{C}{ii}]/CO ratio.
The CO (J=1-0) emission in NGC6824 was not detected (Young et al. 
\cite{Young}).  With the upper limit
the ratio of [\ion{C}{ii}]/CO is estimated to be larger than 10$^4$, which 
is in a similar range 
to those found in quiet spirals (Smith \& Madden \cite{Smith}).
The values of the observed ratio of the other galaxies
$8 \times 10^2 - 3 \times 10^3$ are in agreement with 
those observed in Galactic PDRs (Stacey et al. \cite{Stacey}).
It is slightly smaller than the values predicted in the PDR model.

\begin{figure}
\resizebox{\hsize}{!}{\includegraphics{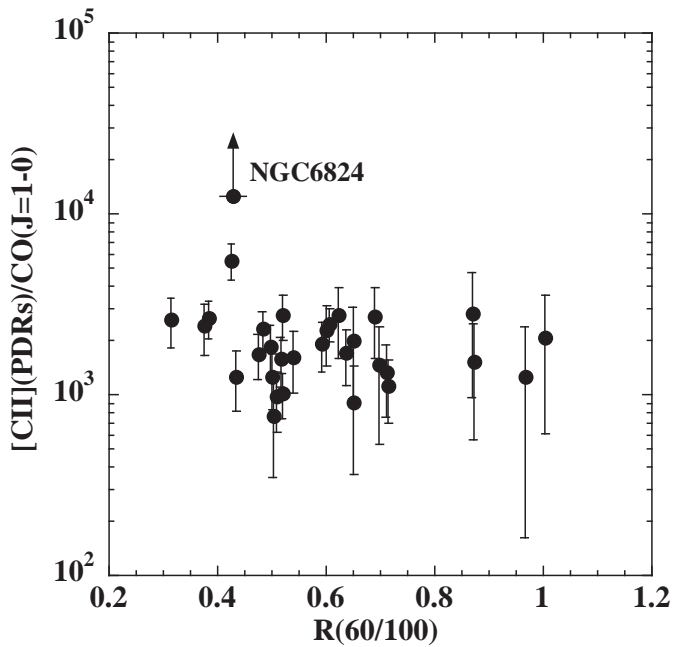}}
\caption{
Ratio of the [\ion{C}{ii}]158\,$\mu$m intensity to the  CO (J=1-0) intensity 
against $R(60/100)$.  The CO emission has not been detected for NGC6824 and a 
lower limit for the ratio is plotted.
}
\label{CO}
\end{figure}

\section{Discussion}
Malhotra et al. (\cite{Malhotra1997}, \cite{Malhotra2001}) have examined six 
possible interpretations for the 
observed decrease in [\ion{C}{ii}]$/FIR$ with the FIR color $R(60/100)$
and favored the interpretation that the charge up of dust 
grains reduces the efficiency of photoelectric heating.
When dust grains are charged positively under strong ultraviolet radiation,
the Coulomb potential prevents electrons from escaping and thus the 
photoelectric heating efficiency decreases.
The electric charge of dust grains is determined by the balance between 
ionization and recombination and $G_\mathrm{0}/n$ is a good measure to 
indicate the electrical potential of dust grains and thus the heating 
efficiency in the parameter range obtained in the 
present study, $G_\mathrm{0} \sim 10^2 - 
10^4$ and $n \sim 10^2 - 10^4 \mathrm{cm}^{-3}$ (Bakes \& Tielens 
\cite{Bakes}; Kaufman et al. \cite{Kaufman}).

To examine this hypothesis, we plot $G_\mathrm{0}/n$ against $R(60/100)$ in 
Fig.~\ref{G0n}.
There is no clear trend seen in the figure, suggesting that at least for the 
present sample of galaxies $G_\mathrm{0}/n$ does not increase with the color. 
The present sample contains starburst galaxies compared to the sample of 
Malhotra et al. (\cite{Malhotra2001}).
The charge up of dust grains is supposed to occur more easily in starbursts 
than in normal galaxies.
However, the decrease of [\ion{C}{ii}]$/FIR$ is more steep in Malhotra 
et al. sample than in the present sample (Fig.~\ref{Malhotra}a).
The interpretation of charge up of dust grains seems to not play a 
significant role in the trend of [\ion{C}{ii}]$/FIR$.  Instead
we suggest that the decrease could be attributed to either the increase
in the de-excitation
of the [\ion{C}{ii}]158\,$\mu$m and/or the decrease in the 
ionized gas component.  The former possibility has been pointed out also by
Genzel \& Cesarsky (\cite{Genzel}).
The latter possibility can be examined by observations of 
[\ion{N}{ii}]205\,$\mu$m which together with the [\ion{N}{ii}]122\,$\mu$m 
intensity allow a better estimate of 
[\ion{C}{ii}]158\,$\mu$m emission from the ionized gas.  The decrease in the 
ionized gas
component may be a consequence of the relative increase in the far-infrared
intensity related to the activities.  Helou et al. (\cite{Helou2001}) have 
also 
suggested that the decrease of the ratio could result from the decrease in the 
importance
of PAHs or small grains in gas heating based on the observed correlation 
between [\ion{C}{ii}]158\,$\mu$m and the $5-10$\,$\mu$m intensities.

\begin{figure}
\resizebox{\hsize}{!}{\includegraphics{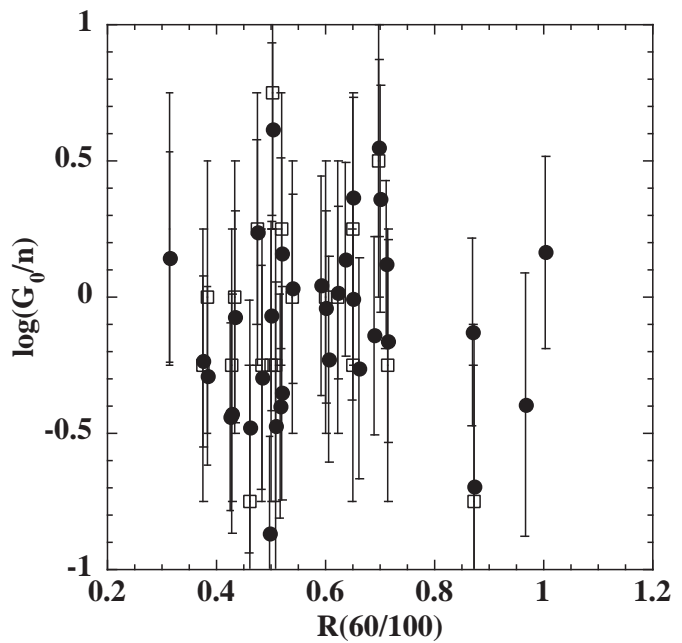}}
\caption{
The ratio of $G_\mathrm{0}$ to $n$ against the 
far-infrared color $R(60/100)$. 
Filled circles indicate the results derived by the comparison of 
[\ion{O}{i}]$/FIR$ with the PDR model, while open squares depict those 
derived by assuming that the [\ion{C}{ii}] from the ionized gas is 
proportional to [\ion{N}{ii}]122\,$\mu$m emission (Eq.~(\ref{conversion})).
}
\label{G0n}
\end{figure}

Based on the observations of H$\alpha$, \ion{H}{i}, CO,
and far-infrared continuum, 
Kennicutt (\cite{Kennicutt}) has
investigated the global Schmidt law in galaxies and found that the 
disk-averaged star formation surface
density $\Sigma_\mathrm{SFR}$ is proportional to $\Sigma_\mathrm{gas}^N$ with 
$N = 1.4 \pm
0.15$, where $\Sigma_\mathrm{gas}$ is the surface density of the gas.
The present results indicate $G_\mathrm{0} \propto n^M$ with $M = 1.25 
^{+0.6}_{-0.3}$
(see Sect.~\ref{sec3}).  If $G_\mathrm{0}$ simply indicates the global 
star-formation rate (SFR)
in galaxies, this relation suggests a similar dependence of SFR on the volume 
gas density of star-forming regions and little variation in the scale height of
star-forming gas distribution in galaxies.  

The present galaxy sample contains AGNs, starburst, and normal galaxies, and 
the present
results are quite similar to those for the normal galaxy sample (Malhotra et 
al. \cite{Malhotra2001}).
Thus they are thought to indicate general characteristics of far-infrared 
properties for a wide 
range of galaxies, though irregular, early-type, and quiescent spiral 
galaxies may show slightly
different characteristics (see  Smith \& Madden \cite{Smith}; Leech et al. 
\cite{Leech}; Malhotra et al. \cite{Malhotra2001a};
Hunter et al. \cite{Hunter}).
We do not see any clear difference in the trends for AGNs in the present
sample, suggesting that the far-infrared properties in AGNs are also
driven by star-forming activities.  A similar conclusion has been drawn
for Cen A by Unger et al. (\cite{Unger}).  ISOPHT observations of 
CfA Seyfert sample by P\'erez Garc\'{\i}a et al. (\cite{Perez}) also indicate
that FIR SED of Seyferts can be interpreted in terms of the thermal emission
from star-forming regions.

We summarize the general trends in the far-infrared properties:
\begin{equation}
[\ion{C}{ii}]/FIR = \left(1.1_{-0.3}^{+0.4} \right) \times 10^{-3} 
R(60/100)^{-\left(0.98_{-0.2}^{+0.5} \right)},
\label{[CII]/FIR_relation}
\end{equation}
\begin{equation}
[\ion{O}{i}]/FIR = (1.25 \pm 0.75) \times 10^{-3},
\label{[OI]/FIR_relation}
\end{equation}
\begin{equation}
[\ion{N}{ii}]/FIR = \left(1.2_{-0.5}^{+0.3} \right) \times 10^{-4} 
R(60/100)^{-\left(1.3_{-0.4}^{+0.7} \right)},
\label{[NII]/FIR_relation}
\end{equation}
and
\begin{equation}
[\ion{O}{iii}]/FIR = \left(1.1_{-0.8}^{+2.0} \right) \times 10^{-3} 
R(60/100)^{\left(1.26_{-0.75}^{+0.25} \right)}.
\label{[OIII]/FIR_relation}
\end{equation}
The last relation of [\ion{O}{iii}]$/FIR$ has a large scatter and should be 
taken with caution.

The present analysis also suggests a general relation between the typical 
density in the
galaxy and the far-infrared color as
\begin{equation}
n = \left(3.0_{-1.5}^{+2.0} \right) \times 10^3 R(60/100)^{(2.4 \pm 0.6)}.
\label{n_color}
\end{equation}

These relations suggest a possibility that the physical properties of 
distant galaxies can be estimated solely from the
far-infrared color.  They can also be used in the comparison of the 
physical properties of distant galaxies to examine differences from nearby
galaxies
in the activities, if any, in observations by future space missions, such as 
SIRTF, ASTRO-F, and Herschel Space Observatory.

\section{Summary}
We have investigated the LWS spectra of $43-197$\,$\mu$m for 34 nearby 
galaxies.
In addition to the detected emission line intensities, we estimated the 
far-infrared
intensity from submicron grains from the continuum emission of $\ge 
80$\,$\mu$m
and derived the average dust temperature.  We obtained the following results.

\begin{enumerate}

\item We have found that the ratios of [\ion{C}{ii}]158\,$\mu$m and 
[\ion{N}{ii}]122\,$\mu$m flux to 
the total far-infrared flux $FIR$ decrease as the far-infrared color
$R(60/100)$ becomes bluer, but the ratio of [\ion{O}{i}]63\,$\mu$m   
to $FIR$ does not show a systematic trend with the color.   The ratio of
[\ion{O}{iii}]88\,$\mu$m to $FIR$ shows a large scatter with a weak trend
of increase with the color.
These are similar to those obtained for the normal galaxy sample 
by Malhotra et al. (\cite{Malhotra2001}).  

\item There is no clear difference seen between AGNs and starburst galaxies 
in the present sample in these trends,
suggesting that even in AGNs the far-infrared properties are dominated by
star-formation activities.

\item Based on the comparison with the PDR model by Kaufman et al. 
(\cite{Kaufman}), we found that the typical neutral gas density
in galaxies increases linearly with the radiation field intensity.

\item About a half of the observed [\ion{C}{ii}]158\,$\mu$m emission is 
estimated to come from PDRs.  The rest can be attributed to that coming from 
the diffuse ionized gas that emits the [\ion{N}{ii}]122\,$\mu$m line.

\item The observed decrease in [\ion{C}{ii}]158\,$\mu$m$/FIR$ with the FIR 
color can be interpreted in terms of either the increase in the collisional 
de-excitation in the [\ion{C}{ii}] transition at high density or the 
decrease 
in the ionized component.  Decrease in the photoelectric yield due to the 
charge up 
of dust grains does not seem to play a primary role in the observed trend.

\item We summarize the relations among the far-infrared properties for the 
present sample
galaxies.  These are thought to indicate the general characteristics for a 
wide range of galaxies,
including starburst, AGNs, and normal galaxies and can be used as a measure 
in the
investigation of activities in distant galaxies in future observations.

\end{enumerate}

\begin{acknowledgements}
The authors thank all the members of Japanese ISO group, particularly H. 
Okuda, K. Kawara, and Y. Satoh for their continuous help and encouragement
and Y. Okada for her help in the calculation of the line ratios.
K.W.C. was supported by the JSPS Postdoctoral Fellowship for Foreign
Researchers.  This work was supported in part by Grant-in-Aids for
Scientific Research from the JSPS.

\end{acknowledgements}

\end{document}